\documentclass[useAMS,usenatbib]{mn2e}
\usepackage{epsfig}
\usepackage{color} 
\usepackage{natbib}
\title[Intranight optical variability of radio-quiet WLQs]
{Intranight Optical Variability of Radio-Quiet Weak Emission Line Quasars}
\author[Gopal-Krishna, Ravi Joshi and Hum Chand]
{{Gopal-Krishna$^{1}$\thanks{E-mail:
krishna@ncra.tifr.res.in (G-K); ravi@aries.res.in (RJ); hum@aries.res.in(HC)},
Ravi Joshi$^{2}$, Hum Chand$^{2}$} \\ 
$^{1}$National Centre for Radio Astrophysics (TIFR), Pune University 
  campus, Pune-411007, India,\\ $^{2}$Aryabhatta Research Institute of 
Observational Sciences (ARIES), Manora Peak, Nainital, 263129, India}

\begin{document}
\date{Accepted ---. Received ---; in original form ---}

\pagerange{\pageref{firstpage}--\pageref{lastpage}} \pubyear{2012}

\maketitle

\label{firstpage}
\begin{abstract}

Based on  a recently started programme, we report the first search 
for intranight optical variability among radio-quiet `weak-line-quasars' 
(RQWLQs). Eight members of this class were observed on 13 nights in 
the $R$-band, such that each source was  monitored continuously
at least once for a minimum duration of about 3.5 hours, using 
the recently installed 130 cm telescope at Devasthal, India. 
Statistical analysis of the differential light curves was carried out
using two versions of the \emph{F$-$}test. Based on the INOV data 
acquired so far, the radio-quiet WLQ population appears to exhibit 
stronger INOV activity as compared to the general population of 
radio-quiet quasars (RQQs), but similar to the INOV known for radio-loud 
quasars of non-blazar type. To improve upon this early result, as well 
as extend the comparison to blazars, a factor of $\sim$2 improvement in 
the INOV detection threshold would be needed. Such efforts are 
underway, motivated by the objective to search for the elusive 
radio-quiet blazars using INOV observations.

\end{abstract}


\section{Introduction} 
\label{sec:intro}
Powerful active galactic nuclei (AGN) whose luminosity across the 
electromagnetic spectrum is dominated by a Doppler boosted relativistic 
jet of nonthermal emission are termed as blazars. The two subsets of
this class, namely BL Lac objects (BLOs) and Highly-Polarized-Quasars 
(HPQs), although differentiated by the equivalent widths of emission 
lines, share many properties. But, whereas HPQs have an
abundant population of weakly polarised quasar counterparts (mostly 
radio-quiet quasars, called RQQs), various searches for radio-quiet analogs 
of BLOs have so far remained unsuccessful. BLOs characterized by very 
weak or absent optical/UV emission lines, which have been pursued in 
such searches, are selected from optical surveys
\citep[e.g.,][]{Jannuzi1993ApJ...404..100J,
  Londish2004MNRAS.352..903L} , although X-ray selected BLOs have also
been targeted \citep[e.g.,][]{Stocke1990ApJ...348..141S}.
Usually, the radio-loudness is quantified in terms of a parameter R
defined as the ratio of the rest-frame 6 cm to 2500\AA ~\rm flux
densities and powerful AGN having $R > 10$ are designated as
radio-loud \citep[e.g.,][]{Jiang2007ApJ...656..680J,
  shen2011ApJS..194...45S,Stocke1992ApJ...396..487S,
  Kellermann1989AJ.....98.1195K}. The first radio-quiet AGN showing
weak emission lines, to be interpreted as a non-BLO
was PG 1407+265 at $z$ = 0.94, based on the lack of variability on 10
year baseline and the lack of optical polarization 
\citep{McDowell1995ApJ...450..585M, 1990ApJS...74..869B}. 
Another example of similar spectral peculiarity is the high accretion-rate
quasar PHL 1811 at $z$ = 0.19 \citep{Leighly2007ApJS..173....1L}.
Samples of radio-quiet BLO candidates at lower redshifts ($z <$ 2.2)
were found in the SDSS survey \citep{York2000AJ....120.1579Y}, by
\citet{Collinge2005AJ....129.2542C} 
and \citet{Anderson2007AJ....133..313A}, and were termed
`Weak-Line-Quasars' (WLQs). As a result, dozens of WLQs marked by 
abnormally weak broad emission-lines \citep[i.e, rest-frame EW
  $<$ 15.4\AA ~\rm for the Ly$\alpha$+NV emission-line
  complex,][]{Diamond-Stanic2009ApJ...699..782D} have been reported
\citep[e.g.,][]{Fan1999ApJ...526L..57F, Fan2006AJ....131.1203F, Anderson2001AJ....122..503A, Collinge2005AJ....129.2542C,
  Schneider2005AJ....130..367S, Schneider2007AJ....134..102S,
  Shemmer2006ApJ...644...86S,Shemmer2009ApJ...696..580S,
  Diamond-Stanic2009ApJ...699..782D, Plotkin2010ApJ...721..562P,
  Plotkin2010AJ....139..390P, Wu2011ApJ...736...28W,
  Hall2002ApJS..141..267H, Hall2004AJ....127.3146H,
  Reimers2005A&A...435...17R, Ganguly2007AJ....133..479G,
  Leighly2007ApJ...663..103L, Hryniewicz2010MNRAS.404.2028H}

Although the above studies have revealed many WLQs that are indeed
radio-quiet \citep[e.g.,][]{Plotkin2010ApJ...721..562P}, they are
commonly identified not as BLOs but RQQs having abnormally weak emission 
lines. This is because, in contrast to BLOs (and much like RQQs), 
the radio-quiet WLQs (RQWLQs) are found to exhibit low optical polarization
\citep{Smith2007ApJ...663..118S} and mild optical continuum
variability on time scales ranging from days to years
\citep{Plotkin2010ApJ...721..562P}. This is further supported by the
similarity observed between the UV-optical spectral indices, $\alpha$,
of WLQs and RQQs. For RQQs the median value of $\alpha$ is -0.52
\citep{Diamond-Stanic2009ApJ...699..782D,
  Plotkin2010AJ....139..390P}, as against -1.15 for BLO candidates 
\citep[e.g.,][]{Plotkin2010AJ....139..390P}. The reason for the
abnormally weak line emission in WLQs is yet to be fully understood, 
but the explanations proposed basically fall into two categories. 
One possible cause of the abnormality is the high mass of the central 
BH (M$_{BH} >~ 3.10^9 M_{\odot}$) which can result in an accretion 
disk too cold to emit strongly the ionizing UV photons, even when its 
optical output is high (\citealt{Laor2011MNRAS.417..681L}; also,
\citealt{Plotkin2010AJ....139..390P}). Alternatively, the covering
factor of the broad-line region (BLR) in WLQs could be at least an
order-of-magnitude smaller compared to the normal QSOs
\citep[e.g.,][]{Nikolajuk2012MNRAS.420.2518N}. An extreme version of
this scenario is that in WLQs the accretion disk is relatively
recently established and hence a significant BLR is yet to develop
\citep{Hryniewicz2010MNRAS.404.2028H, Liu2011ApJ...728L..44L}.
Conceivably, a poor BLR could also result from the weakness of the
radiation pressure driven wind when the AGN is operating at an
exceptionally low accretion rate ($<~ 10^{-2} \ \ to ~\ 10^{-3}
\dot{M}_{Edd}$) (\citealt{Nicastro2003ApJ...589L..13N} ; also,
\citealt{Elitzur2009ApJ...701L..91E}).


While the above mentioned limited empirical evidences and theoretical
scenarios are consistent with the quasar interpretation of the bulk of 
the WLQ population, they do not rule out the possibility of a small 
subset of the population being, in fact, the long-sought radio-quiet 
BLOs in which optical emission arises predominantly from a relativistic 
jet of synchrotron radiation (e.g., \citealt{Stocke1981ApJ...245..375S}, 
\citealt{Diamond-Stanic2009ApJ...699..782D}, 
\citealt{Plotkin2010AJ....139..390P} and references therein; also,
\citealt{Stalin2005MNRAS.359.1022S}).

One strategy to pursue such a search is to characterize the
intra-night optical variability (INOV) of radio-quiet WLQs (RQWLQs). 
It is well established
that normal BLOs (which are always radio-loud) exhibit a distinctly
stronger INOV, both in amplitude ($\psi$) and duty cycle (DC), as
compared to quasars, specially their radio-quiet majority, RQQs
\citep[e.g.,][]{GopalKrishna2003ApJ...586L..25G, 2004MNRAS.350..175S, 
Gupta2005A&A...440..855G, Carini2007AJ....133..303C, Goyal2012A&A...544A..37G}. From this it is evident that INOV properties can be a strong discriminator 
between blazars and other powerful AGN, both radio-loud and radio-quiet
\citep[e.g,][]{2004MNRAS.350..175S,Goyal2012A&A...544A..37G}.
The impetus behind our new programme, therefore, is to characterize 
the INOV behavior of RQWLQs and the first results are presented here.

\section{The sample of radio-quiet WLQs} 
\label{sec:sample}
Our sample for INOV monitoring (Table ~\ref{tab:source_info}) was
derived from the list of 86 radio-quiet WLQs published in Table 6 of
\citep{Plotkin2010AJ....139..390P}, based on the SDSS Data Release 7
\citep[DR-7,][]{Abazajian2009ApJS..182..543A}. Out of that list, we
included in our sample all 18 objects brighter than R $\sim$ 18.5
which are classified as `high-confidence BL Lac candidate'. Thus far,
we have been able to carry out intranight monitoring of only 8 of these 
sources in 13 sessions and the results are reported here.

\begin{table} 
\centering 
\caption{The 8 RQWLQs studied in the present work. \label{tab:source_info}}
\scriptsize 
\begin{tabular}{ccc cc}\\
\hline 
IAU name &  R.A.(J2000) & Dec(J2000)                              &{\it B} &  $z$  \\ 
         & (h m s)      &($ ^{\circ}$ $ ^{\prime}$ $ ^{\prime\prime }$) & (mag)  &       \\
  (1)    &  (2)         & (3)                                     & (4)    & (5)   \\ 
\hline   
\multicolumn{5}{l}{}\\
J081250.79$+$522531.05 &   08 12 50.80& $+$52 25 31 & 18.30 &1.152 \\
J084424.20$+$124546.00 &   08 44 24.20& $+$12 45 46 & 18.28 &2.466 \\
J090107.60$+$384659.00 &   09 01 07.60& $+$38 46 59 & 18.21 &1.329 \\
J121929.50$+$471522.00 &   12 19 29.50& $+$47 15 22 & 17.66 &1.336 \\ 
J125219.50$+$264053.00 &   12 52 19.50& $+$26 40 53 & 17.94 &1.292 \\ 
J142943.60$+$385932.00 &   14 29 43.60& $+$38 59 32 & 17.56 &0.925 \\ 
J153044.10$+$231014.00 &   15 30 44.10& $+$23 10 14 & 17.32 &1.040 \\ 
J161245.68$+$511817.31 &   16 12 45.68& $+$51 18 17 & 17.70 &1.595 \\ 
\hline
\end{tabular}
\end{table}

 \subsection{Photometric observations}

Continuous monitoring of each RQWLQ was done, mainly using the 1.3-m
optical telescope (hereafter 1.3-m DFOT\footnote {Devsthal Fast
  Optical Telescope}) of the Aryabhatta Research Institute of
observational sciencES (ARIES), located at Devasthal, India 
\citep{Sagar2011Csi...101...8.25}.
DFOT is a fast beam (f/4) optical telescope with a
pointing accuracy better than 10 arcsec RMS. The telescope is equipped
with Andor CCD having 2048 $\times$ 2048 pixels of 13.5 micron size, 
resulting in field of view of 18 arcmin on the
sky. The CCD is read out with 31 and 1000 kHz speeds, with the
corresponding system RMS noise of 2.5, 7 e$^{-}$ and gain of 0.7, 2
e${^-}$/Analog-to-Digital Unit (ADU). The camera is cooled down
thermoelectrically to $-$85 degC. We performed continuous monitoring
of each source for about 4 hour in the SDSS$-$r passband at which our
CCD system has maximum sensitivity. For achieving SNR greater than
25-30 our typical exposure time was set between 5$-$8 minutes. The
typical seeing FWHM during our monitoring sessions was 2 arcsec, 
adequate for these point-like sources.

\par

One of the RQWLQ (J125219.47+264053.9) was also monitored with the 1.04-m
Sampurnanand telescope (ST) located at ARIES, Nainital (India).
Another RQWLQ (J090107.60$+$384659.0) was also monitored using the 2-m 
IUCAA Girawali Observatory (IGO) telescope located near Pune (India).
The ST has Ritchey-Chr\'etien (RC) optics with a f$/$13 beam
(Sagar 1999). The detector was a cryogenically cooled 2048 $\times$
2048 chip mounted at the Cassegrain focus. This chip has a readout
noise of 5.3 e$^{-}$/pixel and a gain of 10 e$^{-}$$/$ADU in the slow
readout mode. Each pixel has a area of 24 $\mu$m$^{2}$ which
corresponds to 0.37 arcsec$^{2}$ on the sky, covering a total field of
13$^{\prime}$ $\times$ 13$^{\prime}$. Our observations were carried
out in 2 $\times$ 2 binned mode to improve the signal-to-noise ratio,
and Cousins {\it R} filters were used. \par

The 2-metre IGO telescope has an RC design with a f$/$10 beam at the 
Cassegrain focus\footnote
{http://www.iucaa.ernet.in/$\sim$itp/igoweb/igo$_{-}$tele$_{-}$and$_{-}$inst.htm}.
The detector was a cryogenically cooled 2110$\times$2048 chip mounted at 
the Cassegrain focus. The pixel area is 15 $\mu$m$^{2}$, so that 
the image scale of 0.27 arcsec$/$ pixel covers an area of 10$^{\prime}$
$\times$ 10${^\prime}$ on the sky. The readout noise of this CCD is
4.0 e$^{-}$/pixel and the gain is 1.5 e$^{-}$$/$ADU. The CCD was used
in an unbinned mode with Cousins {\it R} filters. \par

In our sample selection, care was taken to ensure the availability of 
at least two, but usually more, comparison stars on the CCD frame that 
were within about 1 mag of the target RQWLQ. This allowed us to
identify and discount any comparison star which itself varied during a
given night and hence ensured reliable differential photometry of the
RQWLQ.

\subsection{Data Reduction}

All pre-processing of the images (bias subtraction, flat-fielding and
cosmic-ray removal) was carried out using the standard tasks
available in the data reduction software {\textsc
  IRAF} \footnote{\textsc {Image Reduction and Analysis Facility
    (http://iraf.noao.edu/) }}. Instrumental magnitudes of the
comparison stars and the target source were measured from the frames 
using the Dominion Astronomical Observatory Photometry (\textsc{DAOPHOT II})
\textrm{II}\footnote{\textsc {Dominion Astrophysical Observatory
    Photometry}} software designed for concentric circular aperture
photometric technique \citep{1992ASPC...25..297S, 1987PASP...99..191S}. 
As a check on the possible
effects of any seeing variations, the aperture photometry was carried out 
with four aperture radii, ~ 1$\times$FWHM, 2$\times$FWHM, 3$\times$FWHM 
and 4$\times$FWHM, where the seeing disk radius (= FWHM/2) for each
CCD frame was determined using 
5 fairly bright stars on the frame. The data reduced using the four
aperture radii were found to be in generally good
agreement. However, the best S/N for the differential light curves 
(DLCs) was nearly always found for aperture radii of $\sim$2$\times$FWHM, 
so we adopted that aperture for our final analysis.\par

To derive the Differential Light Curves (DLCs) of a given target 
RQWLQ, we selected two steady comparison star present within the CCD 
frames, on the basis of their proximity to the target source,
both in location and magnitude. Coordinates of the comparison star pair 
selected for each RQWLQ are given in Table ~\ref{tab_cdq_comp}. The 
$g-r$ color difference for our `quasar-star' and `star-star' pairs 
is always $< 1.5$, with a median value of 0.54 (column 7, 
Table ~\ref{tab_cdq_comp}). Detailed analyses by
\citet{Carini1992AJ....104...15C} and
\citet{2004MNRAS.350..175S} show that color difference of this
magnitude should produce negligible effect on the DLCs as the 
atmospheric attenuation changes during a monitoring session. \par

Since the selected comparison stars are non-varying, as judged from 
the steadiness of their DLCs, any sharp fluctuation over a single temporal
bin was taken to arise due to improper removal of cosmic rays, or some 
unknown instrumental effect,  and such outlier data points (deviating by 
more than 3$\sigma$ from the mean) were removed from the affected DLCs, 
by applying a mean clip algorithm.
In practice, such outliers were quite rare and never exceeded two data 
points for any DLC, as displayed in Figure 1.
Finally, in order to enhance the SNR, without incurring significant 
loss of time resolution, we have taken 3-point box average of each DLC.

\begin{table*} 
\centering 
\caption{Basic parameters and observing dates of the 8 RQWLQs and their
comparison stars. 
\label{tab_cdq_comp}}
\scriptsize
\begin{tabular}{ccc ccc c}\\
\hline 

{IAU Name} &   Date       &   {R.A.(J2000)} & {Dec.(J2000)}                      & {\it g} & {\it r} & {\it g-r} \\  
           &  dd.mm.yy    &   (h m s)       &($^\circ$ $^\prime$ $^{\prime\prime}$)   & (mag)   & (mag)   & (mag)     \\  
{(1)}      & {(2)}        & {(3)}           & {(4)}                              & {(5)}   & {(6)}   & {(7)}     \\
\hline   
\multicolumn{7}{l}{}\\
J081250.79$+$522531.0 &  23.01.2012      &08 12 50.79 &$+$52 25 31.0  &   18.30 &       18.05 &        0.25\\
S1                    &                  &08 12 50.27 &$+$52 26 32.8  &   17.48 &       17.05 &        0.44\\
S2                    &                  &08 12 29.42 &$+$52 20 49.9  &   19.51 &       18.09 &        1.43\\
J084424.24$+$124546.5 &  26.02.2012      &08 44 24.24 &$+$12 45 46.5  &   18.29 &	17.91 &        0.37\\
S1                    &                  &08 44 30.80 &$+$12 41 24.8  &   19.42 &	18.00 &        1.42\\
S2                    &                  &08 44 39.26 &$+$12 44 54.6  &   18.27 &	17.87 &        0.40\\
J090107.64$+$384658.8 &  27.02.2012      &09 01 07.64 &$+$38 46 58.8  &   18.25 &	18.15 &        0.09\\
S1                    &                  &09 01 21.12 &$+$38 42 14.1  &   18.93 &	17.69 &        1.24\\
S2                    &                  &09 00 43.86 &$+$38 51 42.0  &   17.92 &	17.47 &        0.44\\
J090107.64$+$384658.8 &  16.03.2012      &09 01 07.64 &$+$38 46 58.8  &   18.25 &	18.15 &        0.09\\
S1                    &                  &09 01 00.15 &$+$38 47 09.7  &   19.64 &	18.27 &        1.37\\
S2                    &                  &09 00 59.94 &$+$38 47 51.4  &   18.77 &	18.00 &        0.77\\
J121929.45$+$471522.8 &  26.02.2012      &12 19 29.45 &$+$47 15 22.8  &   17.65 &	17.53 &        0.12\\
S1                    &                  &12 19 33.79 &$+$47 17 04.5  &   17.28 &       16.72 &        0.56\\  
S2                    &                  &12 20 11.17 &$+$47 13 09.2  &   17.88 &       16.83 &        1.05\\
J121929.45$+$471522.8 &  27.04.2012      &12 19 29.45 &$+$47 15 22.8  &   17.65 &	17.53 &        0.12\\
S1                    &                  &12 19 57.89 &$+$47 14 56.9  &   18.66 &	17.35 &        1.31\\
S2                    &                  &12 19 02.24 &$+$47 12 18.2  &   18.42 &	17.18 &        1.24\\
J125219.47$+$264053.9 &  25.02.2012      &12 52 19.47 &$+$26 40 53.9  &   17.94 &	17.70 &        0.24\\
S1                    &                  &12 52 37.93 &$+$26 37 47.6  &   17.52 &	16.98 &        0.54\\
S2                    &                  &12 52 14.26 &$+$26 39 11.5  &   18.43 &	17.15 &        1.28\\
J125219.47$+$264053.9 &  23.03.2012      &12 52 19.47 &$+$26 40 53.9  &   17.94 &	17.70 &        0.24\\
S1                    &                  &12 52 37.93 &$+$26 37 47.6  &   17.52 &	16.98 &        0.54\\
S2                    &                  &12 52 14.26 &$+$26 39 11.5  &   18.43 &	17.15 &        1.28\\
J125219.47$+$264053.9 &  19.05.2012      &12 52 19.47 &$+$26 40 53.9  &   17.94 &	17.70 &        0.24\\
S1                    &                  &12 52 23.82 &$+$26 41 42.6  &   16.71 &	16.42 &        0.29\\
S2                    &                  &12 52 00.81 &$+$26 43 17.5  &   16.93 &	15.86 &        1.07\\
J142943.64$+$385932.2 &  27.02.2012      &14 29 43.64 &$+$38 59 32.2  &   17.56 &	17.55 &        0.01\\
S1                    &                  &14 30 00.65 &$+$38 57 21.4  &   19.08 &	17.62 &        1.46\\
S2                    &                  &14 29 30.69 &$+$39 01 14.2  &   18.16 &	17.00 &        1.16\\
J153044.08$+$231013.4 &  27.04.2012      &15 30 44.08 &$+$23 10 13.4  &   17.83 &	17.59 &        0.24\\
S1                    &                  &15 30 09.51 &$+$23 11 52.9  &   18.46 &	17.15 &        1.31\\
S2                    &                  &15 30 47.76 &$+$23 06 10.4  &   17.79 &	17.19 &        0.60\\
J153044.08$+$231013.4 &  19.05.2012      &15 30 44.08 &$+$23 10 13.4  &   17.83 &	17.59 &        0.24\\            
S1                    &                  &15 30 09.46 &$+$23 11 07.1  &   17.29 &	16.71 &        0.58\\
S2                    &                  &15 30 57.70 &$+$23 07 42.3  &   17.01 &	16.65 &        0.36\\
J161245.68$+$511816.9 &  18.05.2012      &16 12 45.68 &$+$51 18 16.9  &   17.89 &	17.72 &        0.17\\
S1                    &                  &16 12 26.15 &$+$51 22 14.6  &   18.08 &	16.69 &        1.39\\
S2                    &                  &16 12 48.21 &$+$51 18 37.1  &   15.33 &	14.94 &        0.39\\ 
\hline
\end{tabular}
\end{table*}

\section{Analysis}
Conventionally, the presence of INOV in a DLC is quantified using
$C$-statistics \citep{1997AJ....114..565J}. However, recently
\citet{Diego2010AJ....139.1269D} has pointed out that this is not a
valid test as it is based on ratio of two standard deviations which
(unlike variance) are not linear operators and the nominal critical
value used for confirming the presence of variability (i.e., 2.576) 
is usually too conservative. He has therefore advocated more powerful 
statistical tests, namely, the one-way analysis of variance (ANOVA) 
and the \emph{F$-$}test. However, a proper use of the ANOVA test 
requires a rather large number of data points in the DLC, so as to 
have several points within each subgroup used for the analysis; this 
is not feasible for our light curves which typically have only around 
15 - 20 data points each. Therefore, in this study we shall rely on 
the \emph{F$-$}test which is based on the ratio of variances as,
$F={variance_{(observed)}}/{variance_{(expected)}}$
(\citet{Diego2010AJ....139.1269D}. Two versions of this test
employed in the recent literature are: (i) the standard \emph{F$-$}test
(hereinafter, $F^{\eta}-$test, \citet{Goyal2012A&A...544A..37G} and
(ii) scaled \emph{F$-$}test, hereinafter, $F^{\kappa}-$test
\citet{Joshi2011MNRAS.412.2717J}. In this work we have subjected all
our DLCs to both these statistical tests, as discussed below.

An important point to be borne in mind while applying the
$F^{\eta}-$test is that the photometric errors, as returned by the
routines in the \textsc{IRAF}  and \textsc{DAOPHOT} softwares, are normally underestimated 
by a factor $\eta$ ranging between 
$1.3$ and $1.75$, as found in independent studies \citep[e.g.,][]
{1995MNRAS.274..701G, 1999MNRAS.309..803G, Sagar2004MNRAS.348..176S, 
Stalin2004JApA...25....1S, Bachev2005MNRAS.358..774B}.
In a recent analysis of 73 DLCs derived for 73 pairs of `steady' stars 
monitored on as many nights, \citet{Goyal2012A&A...544A..37G} estimated 
the best-fit value of $\eta$ to be 1.5.
(see, also Sect. 4). The $F^{\eta}-$ statistics can be expressed as:

\begin{equation}
 \label{eq.fetest}
F_{1}^{\eta} = \frac{\sigma^{2}_{(q-s1)}}
{ \eta^2 \langle \sigma_{q-s1}^2 \rangle}, \nonumber  \\
\hspace{0.2cm} F_{2}^{\eta} = \frac{\sigma^{2}_{(q-s2)}}
{ \eta^2 \langle \sigma_{q-s2}^2 \rangle},\nonumber  \\
\hspace{0.2cm} F_{s1-s2}^{\eta} = \frac{\sigma^{2}_{(s1-s2)}}
{ \eta^2 \langle \sigma_{s1-s2}^2 \rangle}.
\end{equation}


where $\sigma^{2}_{(q-s1)}$, $\sigma^{2}_{(q-s2)}$ and
$\sigma^{2}_{(s1-s2)}$ are the variances of the `quasar-star1',
`quasar-star2' and `star1-star2' DLCs and $\langle \sigma_{q-s1}^2
\rangle=\sum_\mathbf{i=0}^{N}\sigma^2_{i,err}(q-s1)/N$, $\langle \sigma_{q-s2}^2 \rangle$ and 
$\langle \sigma_{s1-s2}^2 \rangle$ are the mean square (formal) rms
errors of the individual data points in the `quasar-star1', `quasar-star2' 
and `star1-star2' DLCs, respectively. $\eta$ is the scaling factor 
\citep[$= 1.5$, cf.][]{Goyal2012A&A...544A..37G}, introduced to account 
for the underestimation of photometric rms errors returned by the photometry 
algorithms used here, as mentioned above.

The $F$ values computed using Eq. ~\ref{eq.fetest} , were then compared
individually with the critical $F$ value,
$F^{(\alpha)}_{\nu_{qs},\nu_{ss}}$, where $\alpha$ is the significance
level set for the test, and $\nu_{qs}$ and $\nu_{ss}$ are the degrees
of freedom of the `quasar-star' and `star-star' DLCs, respectively. The
smaller the $\alpha$, the more improbable is the result to arise
from chance. For the present study, we have used two significance
levels, $\alpha=$ 0.01 and 0.05, which correspond to confidence levels
of greater than 99 and 95 per cent, respectively. If $F$ is found to
exceed the critical value, the null hypothesis (i.e., no variability)
is discarded to the corresponding level of confidence. We have computed 
separately the \emph{F$-$}values for the `quasar-star1' and `quasar-star2' 
DLCs (i.e., $F_1^\eta$ \& $F_2^\eta$) from Eq.~\ref{eq.fetest}.
Thus, for a given monitoring session, a RQWLQ is marked as \emph{variable} 
(`V') if for both its DLCs $F$-value $\ge F_{c}(0.99)$, which corresponds 
to a confidence level $\ge 99$ per cent; \emph{non-variable} (`NV') if 
even one of the two DLCs is found to have $F$-value less than 
$F_{c}(0.95)$. The remaining cases are termed as \emph{probably variable} 
(`PV').

An alternative approach to quantify the INOV status of a DLC has been
followed in \citet{Joshi2011MNRAS.412.2717J}, the `scaled \emph{F$-$}test'.
Instead of $\eta$, this test relies on a factor $\kappa$ equal to the ratio 
of the mean square rms errors of the data points in the quasar DLC 
relative to a comparison star and in the DLC of that star relative
to the other comparison star. 
This parameter is intended to correct for any bias which may arise 
due to some systematic difference between the photometric errors 
of the data points in the `quasar-star' and `star-star' DLCs (e.g., 
due to a brightness mismatch between the quasar and the comparison star(s)). 
Thus, in this `scaled' \emph{F$-$}test,

\begin{equation}
 F_{1}^{\kappa}=\frac{var(q-s1)}{\kappa \times var(s1-s2)},  \\
 F_{2}^{\kappa}=\frac{var(q-s2)}{\kappa \times var(s1-s2)}
\label{eq.fstest}
\end{equation}

with  $\kappa$, defined as,

\begin{equation}
\kappa=\left[\displaystyle{\frac{\sum_\mathbf{i=0}^{N}\sigma^2_{i,err}(q-s)/N}{\sum_{i=0}^{N}\sigma^2_{i,err}(s1-s2)/N}}\right] \equiv \frac{\langle\sigma^2_{q-s}\rangle}{\langle\sigma^2_{s1-s2}\rangle},
\label{eq:kappa}
\end{equation}
where $\sigma_{i,err}(q-s)$ and $\sigma_{i,err}(s1-s2)$ are,
respectively, the rms errors on individual points of the `quasar-star' and
`star-star' DLCs, as returned by the \textsc{DAOPHOT/IRAF} routine.

The threshold criteria for inferring the INOV status of a DLC from its  
computed $F$-value in this $F^{\kappa}$-test is identical to that 
adopted above for the $F^{\eta}$-test. The inferred INOV status of the 
DLCs of each RQWLQ, relative to two comparison stars, are presented in
Table~\ref{tab:res}. In the first 5 columns, we list the name of the
RQWLQ, date of its monitoring, telescope used, duration of monitoring
and the number, N, of data points in the DLCs relative to the two comparison 
stars (s1 and s2). The next two columns give the computed $F$-values, 
based on the $F^{\eta}-$test and $F^{\kappa}-$tests. Column 8 and 9 
mention the INOV status of the two DLCs of the RQWLQ, as inferred from 
the $F^{\eta}-$test and $F^{\kappa}-$test, respectively. Column 10 
gives the INOV amplitudes $\psi$ derived from the two DLCs of the RQWLQ, 
based on the definition given by \citet*{Romero1999A&AS..135..477R}:

\begin{equation} 
\psi= \sqrt{({D_{max}}-{D_{min}})^2-2\sigma^2} 
\end{equation} 

with  $D_{min,max}$ = minimum (maximum) in the RQWLQ DLC and $\sigma^2$= 
$\eta^2$$\langle\sigma^2_{q-s}\rangle$, where, 
$\eta$ =1.5 \citep{Goyal2012A&A...544A..37G}.
Column 11 lists the square root of the scaling factor,$\kappa$ (Eq. 3).
which has been used to scale the variance of the star$-$star DLCs while 
computing the $F$-value in the scaled $F$-test (Eq. 2). The last column gives 
our averaged photometric error  $\sigma_{i,err}(q-s)$ in the 
`quasar$-$star' DLCs (i.e., mean value for q-s1 and q-s2 DLCs),
which typically lies between 0.01 and 0.02 mag.

\subsection{The INOV duty cycle (\emph{DC})}

To recapitulate, a RQWLQ in a given session is marked as
\emph{variable} (`V') if its DLCs relative to the two comparison
stars, are both found to have $F$-value $\ge F_{c}(0.99)$, which
corresponds to a confidence level $\ge 99$ per cent; \emph{non-variable}
(`NV') if even one of the two DLCs is found to have $F$-value less
than $F_{c}(0.95)$. The remaining cases are marked as \emph{probably
variable} (`PV').

The duty cycle of INOV was computed using the definition by
\citet*{Romero1999A&AS..135..477R},
\begin{equation} 
DC = 100\frac{\sum_{i=1}^n N_i(1/\Delta t_i)}{\sum_{i=1}^n (1/\Delta t_i)} 
{\rm per cent} 
\label{eqno1} 
\end{equation} 
where $\Delta t_i = \Delta t_{i,obs}(1+z)^{-1}$ is duration of the
monitoring session of a source on the $i^{th}$ night, corrected for
its cosmological redshift, $z$. Since the duration of the observing 
session for a given source differs from night to night,
the computation of DC has been weighted by the actual monitoring 
duration $\Delta t_i$ on the $i^{th}$ night. $N_i$ was set equal to 1, 
if INOV was detected (i.e., `V'), otherwise $N_i$ was taken as zero.

\section{Discussion and Conclusions}

The present study marks the beginning of a systematic investigation of the 
INOV properties of radio-quiet weak-line-quasars (RQWLQs). This initial
attempt is based on a modest size sample containing 8 RQWLQs, for 
which the derived results are presented in Table~3. Using the 
$F^{\kappa}-$test we obtained an INOV duty cycle (\emph{DC}) of 
$\sim$ 13 per cent which rises to $\sim$ 30 per cent if the two 
cases of probable INOV (`PV') are included. On the other hand, the 
$F^{\eta}-$test yields for the same dataset an INOV \emph{DC} 
of $\sim$ 6 per cent (taking the best-fit value of $\eta$ = 1.5, Sect.\ 3). 
Thus, taken together, the two $F-$tests lead to an average INOV \emph{DC} 
of around 9 per cent for RQWLQs, for monitoring sessions lasting $\ga$ 
3.5 hours.  In order to assess the effect of possible 
uncertainty in the $\eta$ factor (Sect.\ 3), we have repeated the 
$F^{\eta}-$test for the entire sample, taking two extreme values of 
$\eta$ (= 1.3 and 1.75), as reported in the literature ~ 
\citep[e.g.,][]{1995MNRAS.274..701G, 1999MNRAS.309..803G, 
2004MNRAS.350..175S, Bachev2005MNRAS.358..774B}. 
The computed INOV duty cycles for both these extreme values of $\eta$ 
are still 6 per cent, i.e., the same as that estimated above taking 
$\eta$ = 1.5, the best-fit estimate given in \citet{Goyal2012A&A...544A..37G}.
Thus, the $F^{\eta}-$test is found to give consistent results over the 
maximum plausible range of uncertainty in $\eta$.

 At this point it seems worthwhile to also mention the DC estimate 
based on the more conservative, but hitherto much more extensively used
$C-$test (Sect. 3).  We find that the only change to Table 3, resulting 
from the application of $C-$test to our dataset is that INOV status of 
the WLQ J121929.45$+$471522.8 on 26.02.2012 changes from `V' to `PV'. 
This leaves no clear incidence of INOV detection in the present data. 
Treating `PV' cases as `V' yields an INOV duty cycle of $\sim 6$ percent, 
which would clearly be an upper limit, albeit using a small sample. Our 
subsequent discussion will only be based on the results obtained from the 
$F-$test as it is believed to be a more powerful test (Diego 2010; Sect. 3).

Bearing in mind the modest size of our RQWLQ sample at this stage,
we now attempt a comparison of the INOV duty cycle with the estimates 
available for RQQs and other AGN classes, such as non-blazar type 
flat-spectrum radio quasars (FSRQs) and blazars.
INOV duty cycles for these AGN classes have been
extensively reported in the literature 
\citep[e.g.,][]{Stalin2004JApA...25....1S, Goyal2012A&A...544A..37G}, 
mostly based on DLCs longer than $\sim$ 4-hours (which broadly holds even for the
present DLCs of RQWLQs, as well). One limitation encountered in 
making the comparison is that for the observations of all these other 
AGN types, an INOV detection threshold ($\psi_{lim}$) of $1\%$ to  $2\%$  
had typically been achieved (at least in our programme from ARIES, 
Sect. ~1).  Being 1 to 2 magnitudes fainter, the INOV detection threshold 
reached for the present sample of RQWLQs is less deep ($\psi_{lim} \sim 
4 - 5\%$, Table 3). Thus, for the purpose of comparison with the 
afore-mentioned other AGN types, our present estimate of INOV DC for 
RQWLQs ($\sim 9$ per cent) must be treated as a lower limit. It would 
be very interesting to check if a factor of $2-3$ improvement in 
$\psi_{lim}$ would lead to a much higher INOV DC for RQWLQs, perhaps 
even approaching the level of $\sim 50\%$ which is established for 
strong INOV (i.e., $\psi > 3\%$) of blazars (BL Lacs and high-polarization 
radio quasars) when they are monitored for $\ga$ 4 hours.
\citep[e.g.,][]{GopalKrishna2003ApJ...586L..25G, 2004MNRAS.350..175S, 
Stalin2004JApA...25....1S, Sagar2004MNRAS.348..176S, 
GopalKrishna2011MNRAS.416..101G, Goyal2012A&A...544A..37G}. 
The DC for strong INOV is found to be only $\sim 7\%$ for non-blazar type
FSRQs (based on the $F^{\eta}-$test,  
\citep[e.g.,][]{Goyal2012A&A...544A..37G} and 
practically zero for radio-quiet quasars 
since they are not known to show INOV with $\psi > 3\%$
 \citep[e.g.,][]{2007BASI...35..141G, 2004MNRAS.350..175S, Stalin2005MNRAS.356..607S, 
GopalKrishna2003ApJ...586L..25G}. Thus, one indication emerging from 
this first INOV observations of radio-quiet WLQs is that their INOV 
level, as a class, is likely to be significantly stronger in comparison 
to the general population of radio-quiet quasars and, indeed similar
that that known for non-blazar type radio quasars (FSRQs). It remains
to be seen whether on attaining a matching INOV detection threshold 
$\psi_{lim} \sim 1 - 2\%$, the INOV activity level of RQWLQs will be 
found to be stronger, perhaps approaching the high levels exhibited by 
blazars (\citep[e.g.,]
[]{Goyal2012A&A...544A..37G} and references therein). 
This remains an outstanding question to be pursued, in view of its 
potential for unravelling the nature of WLQs and for the key question 
whether radio-quiet BL Lacs at all exist (Sect. 1). It may be noted 
here that a hint that, compared to normal RQQs,  RQWLQs may show stronger 
optical/UV variability on {\it year-like} time scale, has been reported by
\citet{Stalin2005MNRAS.359.1022S}; though it is based on monitoring of 
just one RQWLQ (SDSS J153259.96$-$003944.1 at $z$ = 4.67).\par
 
To summarize, the twin objectives pursued in this exploratory, first
INOV study of radio-quiet weak-line-quasars (RQWLQs) are: (a) To find
cases of strong INOV ($\psi > 3\%$), any such RQWLQs would be
outstanding candidates for the putative radio-quiet BL Lacs, and (b)
To quantify the INOV duty cycle for the class of RQWLQs, for both 
strong and weaker INOV. In our program we have so far been able to 
cover only a modest size sample containing 8 RQWLQs, each monitored 
in at least one session lasting $\ga$ 3.5 hours. This has led to the
result that the duty cycle of strong INOV in this class of AGNs 
seems to be higher than that known for radio-quiet quasars and is
similar to that known for (non-blazar) FSRQs. This early indication 
provides impetus to continue this programme, in particular, to check if 
blazar-like INOV levels occur in some RQWLQs. To attain the required
observational capability, a factor of $\ga 2$ improvement in the INOV 
detection threshold would be needed and we are attempting to achieve
this by monitoring relatively bright RQWLQs on dark nights, possibly 
using a telescope larger than the newly installed 1.3-metre DFOT used 
in the present work.

\vskip0.15in

{\bf Acknowledgements}

We would like to thank Dr. Arti Goyal for helpful discussions and
the scientific staff and observers at the 1.3-m DFOT telescope, 
ARIES (Nainital, India) for the assistance with the observations.
       
 \begin{table*}
  \centering
  \begin{minipage}{500mm}
 {\small
 \caption{Observational details and INOV results for the sample of 8 RQWLQs.}
 \label{tab:res}
 \begin{tabular}{@{}ccc cc rrr rrr ccc@{}} 
 \hline  \multicolumn{1}{c}{RQWLQ} &{Date} &{Tel.} &{T} &{N{\footnote{The number of data points after three point box average.}}} 
 &\multicolumn{2}{c}{F-test values}
 &\multicolumn{2}{c}{INOV status{\footnote{V=variable, i.e., confidence level
       $\ge 0.99$; PV=probable variable, i.e., $0.95-0.99$ confidence level; 
       NV=non-variable, i.e., confidence level $< 0.95$.\\
 Variability status values based on quasar-star1 and quasar-star2
 pairs are separated by a comma.}}}
 &INOV amplitude
 &{$\sqrt\kappa${\footnote{Here
        $\kappa=\langle\sigma^2_{q-s}\rangle/\langle\sigma^2_{s1-s2}\rangle$ (as in Eq.~\ref{eq:kappa}), is
  used to scale the variance of star1-star2 DLCs for  the scaled F-test.}}}
 &{$\sqrt { \langle \sigma^2_{i,err} \rangle}$}\\ & dd.mm.yyyy& used& 
hr & & {$F_1^{\eta}$},{$F_2^{\eta}$}&{$F_1^{\kappa}$},{$F_2^{\kappa}$}
 &F$_{\eta}$-test &F$_{\kappa}$-test & $\psi_1(\%),\psi_2$(\%)&$\frac{}{}$ 
&(q-s)\\ (1)&(2) &(3) &(4) &(5) &(6)
 &(7)&(8)&(9) &(10) &(11) &(12)\\ \hline

J081250.79$+$522530.9  &23.01.2012 &  DFOT & 5.70&  13 & 0.77, 0.59  &  1.59,1.21  & NV, NV&   NV, NV &  3.03,1.94  &  0.99 & 0.01\\ \\
J084424.24$+$124546.5  &26.02.2012 &  DFOT & 4.28&  17 & 0.65, 0.63  &  2.83,2.74  & NV, NV&   PV, PV &  4.49,3.49  &  1.00 & 0.01\\ \\
J090107.64$+$384658.8  &27.02.2012 &  DFOT & 3.86&  12 & 1.62, 1.67  &  5.81,6.00  & NV, NV&   V , V  &  5.00,4.74  &  1.41 & 0.01\\   
J090107.64$+$384658.8  &16.03.2012 &  IGO  & 3.52&  07 & 1.11, 0.57  &  2.66,1.36  & NV, NV&   NV, NV &  3.73,2.37  &  1.07 & 0.01\\ \\
J121929.45$+$471522.8  &26.02.2012 &  DFOT & 4.87&  23 & 4.85, 6.23  &  5.55,7.13  & V ,V  &   V ,V   &  6.35,6.14  &  1.52 & 0.01\\   
J121929.45$+$471522.8  &27.04.2012 &  DFOT & 3.02&  15 & 1.01, 1.65  &  1.34,2.18  & NV, NV&   NV, NV &  4.56,6.64  &  1.13 & 0.01\\ \\
J125219.47$+$264053.9  &25.02.2012 &  DFOT & 2.23&  09 & 0.24, 0.37  &  0.21,0.32  & NV, NV&   NV, NV &  0.36,1.39  &  1.43 & 0.01\\   
J125219.47$+$264053.9  &23.03.2012 &  ST   & 3.45&  09 & 0.98, 1.02  &  3.00,3.12  & NV, NV&   NV, NV &  3.93,3.87  &  1.51 & 0.01\\   
J125219.47$+$264053.9  &19.05.2012 &  DFOT & 3.81&  15 & 0.52, 0.54  &  0.92,0.95  & NV, NV&   NV, NV &  3.43,3.76  &  3.17 & 0.01\\ \\
J142943.64$+$385932.2  &27.02.2012 &  DFOT & 3.76&  18 & 0.46, 1.41  &  1.23,3.76  & NV, NV&   NV, V  &  3.49,4.58  &  1.05 & 0.01\\ \\
J153044.07$+$231013.5  &27.04.2012 &  DFOT & 4.07&  20 & 2.13, 1.48  &  3.07,2.13  & NV, NV&   V ,NV  &  5.46,3.81  &  1.21 & 0.01\\   
J153044.07$+$231013.5  &19.05.2012 &  DFOT & 3.21&  13 & 0.67, 0.58  &  3.63,3.12  & NV, NV&   PV, PV &  4.19,4.02  &  1.62 & 0.01\\ \\
J161245.68$+$511817.3  &18.05.2012 &  DFOT & 4.03&  16 & 0.44, 0.44  &  1.81,1.83  & NV, NV&   NV, NV &  4.02,3.87  &  3.57 & 0.02\\   

 \hline
 \end{tabular}  
 }  
 \end{minipage}
 \end{table*}


\begin{figure*}
\centering
\epsfig{figure=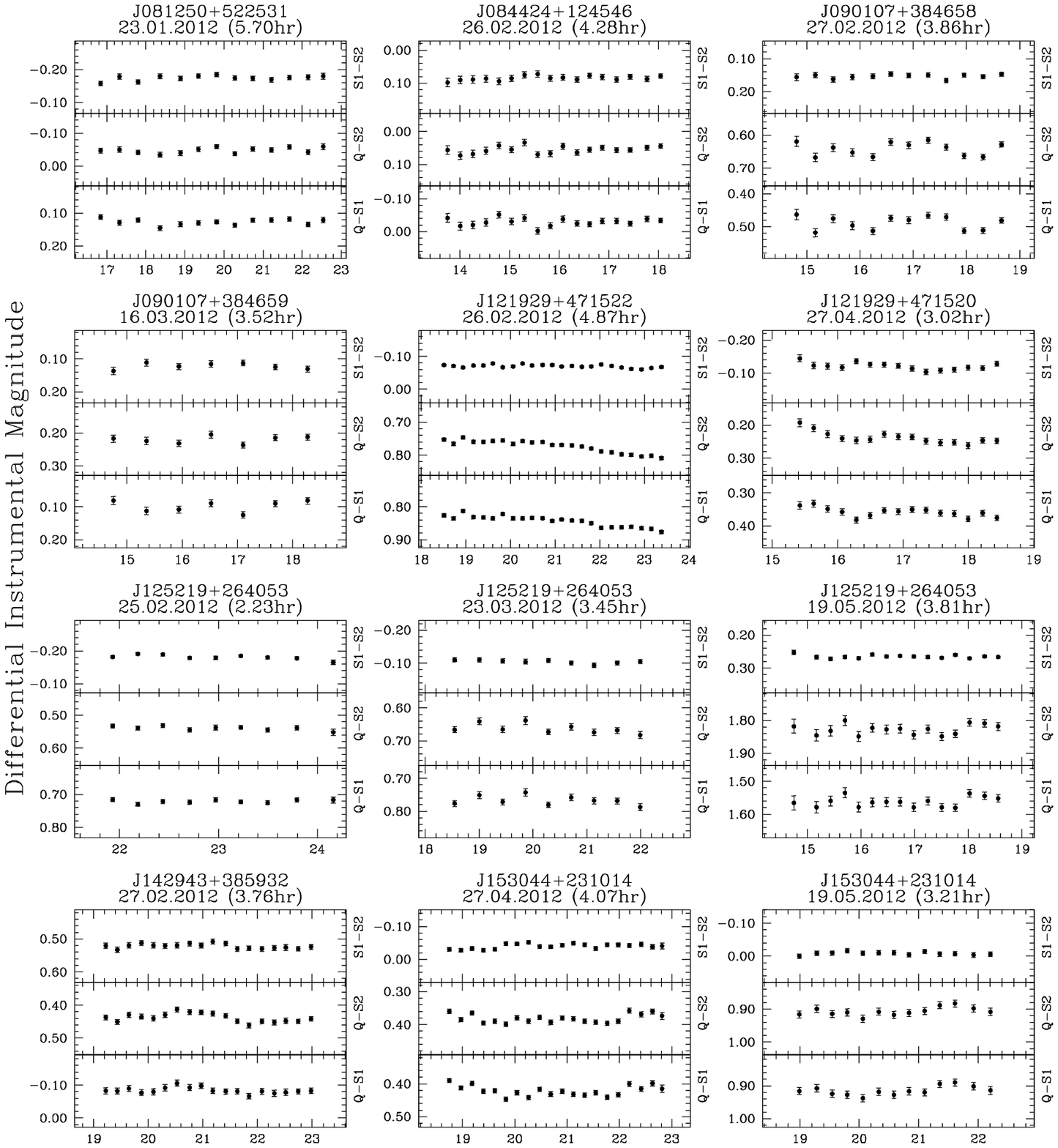,height=17.5cm,angle=00,bbllx=18bp,bblly=143bp,bburx=541bp,bbury=711bp,clip=true}
\epsfig{figure=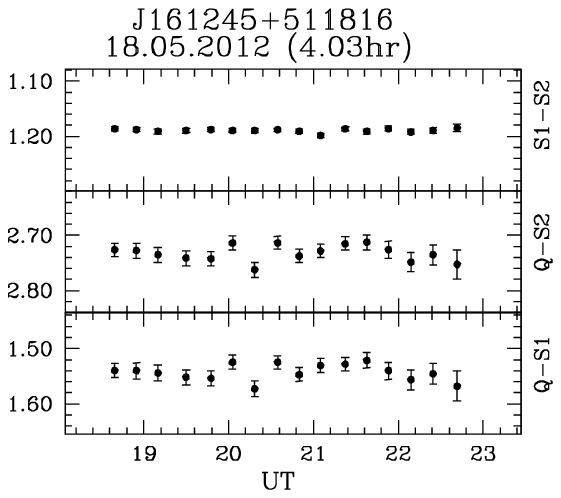,height=4.5cm,width=5cm,angle=00,bbllx=32bp,bblly=145bp,bburx=198bp,bbury=288bp,clip=true}
\caption[]{Differential light curves (DLCs), after three point box average, for the 8 RQWLQs in our sample. The name of the quasar along with the date and duration of the monitoring 
 session are given at the top of each panel. In each panel the upper DLC is 
 derived using the two comparison stars, while the lower two DLCs are the 
 `quasar-star' DLCs, as defined in the labels on the right side. Any likely 
 outlier point (at $> 3\sigma$) in the DLCs are marked with crosses (see
 Sect. 2) and those points are excluded from the statistical analysis.}
\label{fig:lurve}
 \end{figure*}

\label{lastpage}

\bibliography{references}
\end{document}